\documentclass[12pt,a4paper,final]{iopart}

\usepackage{iopams}
\usepackage{graphicx}
\usepackage[breaklinks=true,colorlinks=true,linkcolor=blue,urlcolor=blue,citecolor=blue]{hyperref}
\usepackage[font=footnotesize,labelfont=bf,
   justification=justified,
   format=plain]{caption}

\begin{document}

\title[]{Tuning of Electrical,  Magnetic, and Topological Properties of Magnetic Weyl Semimetal Mn$_{3+x}$Ge by Fe doping}

\author{Susanta Ghosh, Achintya Low, Soumya Ghorai, Kalyan Mandal, and Setti Thirupathaiah$^*$}
\address{Department of Condensed Matter and Materials Physics, S. N. Bose National Centre for Basic Sciences, Kolkata, West Bengal-700106, India}
\eads{\mailto{setti@bose.res.in}}

\begin{indented}
\item[]\today
\end{indented}

\begin{abstract}
We report on the tuning of electrical, magnetic, and topological properties of the magnetic Weyl semimetal (Mn$_{3+x}$Ge) by Fe doping at the Mn site, Mn$_{(3+x)-\delta}$Fe$_{\delta}$Ge ($\delta$=0, 0.30, and 0.62). Fe doping significantly changes the electrical and magnetic properties of Mn$_{3+x}$Ge. The resistivity of the parent compound displays metallic behavior, the system with $\delta$=0.30 of Fe doping exhibits semiconducting or bad-metallic behavior, and the system with $\delta$=0.62 of Fe doping demonstrates a metal-insulator transition at around 100 K. Further, we observe that the Fe doping increases in-plane ferromagnetism, magnetocrystalline anisotropy, and induces a spin-glass state at low temperatures. Surprisingly, topological Hall state has been noticed at a Fe doping of $\delta$=0.30 that is not found in the parent compound or with $\delta$=0.62 of Fe doping. In addition, spontaneous anomalous Hall effect observed in the parent system is significantly reduced with increasing Fe doping concentration.
\end{abstract}

 \vspace{2pc}                             
\section{Introduction}

Topological materials with kagome lattice structure offer many intriguing quantum phenomena such as the Weyl \cite{Xu2011, Liu2019} and Dirac fermions \cite{Liu2020, Ye2018} in solids, quantum spin liquid state \cite{Nakano2017}, superconductivity \cite{Wu2021}, anomalous Hall effect (AHE) \cite{Liu2018}, topological Hall effect (THE) \cite{Li2019}, and skyrmion lattice \cite{Kanazawa2011, Muehlbauer2009}. Though there exist many topological kagome systems, the topological kagome magnets have gained much attention due to their potential technological applications in spintronics as well as from a fundamental science point of view due to strong electronic correlations present in these systems \cite{Zhang2021, Chowdhury2023}. For instance, Co$_3Sn_2S_2$ is a ferromagnetic Weyl semimetal showing giant AHE and zero-field Nernst effect \cite{Liu2018, Morali2019}. The noncollinear and coplanar antiferromagnetic (AFM) Weyl semimetals, Mn$_3X$ (X = Sn and Ge), show large AHE originating from the nonzero Berry phase in momentum space~\cite{Nakatsuji2015, Nayak2016, Kiyohara2016, Rout2019, Taylor2020}. On the other hand, ferromagnetic Weyl semimetal Fe$_3Sn_2$ shows both THE and AHE \cite{Li2019a}. A nonzero Berry phase in the momentum space leads to an intrinsic anomalous Hall effect. In contrast, the nonzero total scalar-spin chirality [$\chi_{ijk}$ = $S_i.(S_j \times S_k)$] summed over the lattice,  producing the nonzero real-space Berry curvature,  generates the topological Hall effect \cite{Shiomi2012, Denisov2018, Rout2019}.

In noncentrosymmetric magnets such as MnSi \cite{Neubauer2009, Nagaosa2013}, FeGe \cite{Yu2010, Huang2012}, MnGe \cite{Kanazawa2011}, and FeCoSi \cite{Yu2010a}, THE originates from the chiral-spin structure stabilized by the Dzyaloshinskii-Moriya interaction (DMI).On the other hand, in centrosymmetric magnets such as La$_{1-x}Sr_xMnO_3$ \cite{Yu2014}, Fe$_3Sn_2$ \cite{Hou2018, Du2022}, and Mn$_{3-\delta}Fe_{\delta}Sn$ \cite{Low2022}, THE originates from the chiral-spin structure stabilized by strong magnetic anisotropy. To date, many systems of various magnetic orderings have been found to show THE, including the skyrmionic crystals \cite{Neubauer2009, Spencer2018, Li2013}, the antiferromagnets (AFM) \cite{Suergers2014, Ueland2012}, the spin glass systems \cite{Taniguchi2004, Fabris2006}, the frustrated magnets \cite{Machida2007, Takatsu2010}, the double-exchanged ferromagnets \cite{Wang2006, Vistoli2018}, and the magnetic skyrmion arrays \cite{Ohuchi2015, Soumyanarayanan2017, Liu2017, Kanazawa2015}. Among them, the frustrated kagome magnets such as PdCrO$_2$ \cite{Takatsu2010}, $Gd_2PdSi_3$ \cite{ Kurumaji2019}, and $Mn_{3+x}$Sn and $Fe_3Sn_2$ \cite{Li2019, Rout2019}  are more interesting as they are rich in physics due to the triangular spin-lattice in the kagome network. Mn$_3$Ge, like its sister compound Mn$_3$Sn, is a noncollinear antiferromagnet with a hexagonal crystal structure of space group $P6_3/mmc$ \cite{Chen2021, Nakatsuji2015, Kiyohara2016, Nayak2016}. Here, the Mn atoms form the kagome network with triangles and hexagons in the $ab$-plane of the crystal, with the Ge atom sitting at the centre of hexagon \cite{Yamada1988, Arras2011, Nayak2016, Kiyohara2016}. Mn$_{3+x}Sn$ shows a spin-reorientation transition at around 260–270 K, while no such spin-reorientation transition is found in Mn$_{3+x}Ge$ down to the lowest possible temperature. Further, Mn$_{3+x}Sn$ shows a low-temperature spin-glass transition at around 50 K but is not found in Mn$_{3+x}Ge$. Earlier, we found that Fe doping at the Mn site can significantly tune the electrical, magnetic, and magnetotransport properties of Mn$_{3-\delta}Fe_{\delta}Sn$ \cite{Low2022}.

In this contribution, we systematically studied the electrical, magnetic, and Hall effect properties of Mn$_{(3+x)-\delta}Fe_{\delta}Ge$ ($\delta$=0, 0.30, and 0.62) by varying the Fe doping concentration. Fe doping significantly changes the electrical resistivity and magnetic properties of Mn$_{3+x}$Ge. Most importantly, we observe a topological Hall state only in the compound with $\delta$=0.30 of Fe doping but not in the parent system and the system with $\delta$=0.62 of Fe doping. In addition, increasing the Fe doping concentration significantly reduces the spontaneous anomalous Hall effect observed in the parent system. In the manuscript, we thoroughly explain our experimental observations.

\begin{table*}
\caption{Lattice parameters obtained from the single crystal XRD (SCXRD) measurements.}
  \centering
\begin{tabular*}{\linewidth}{c @{\extracolsep{\fill}} ccccccc}
 \hline\hline\\
Composition & a ($\AA$) & b ($\AA$) & c ($\AA$) & $\alpha$ ($^o$)  & $\beta$ ($^o$)  & $\gamma$ ($^o$)\\[1ex]
 \hline\hline\\[1ex]
Mn$_{3.48}$Ge ($\delta$=0) & 5.350(5)  & 5.352(5) & 4.308(9)  & 90.06 (17)& 90.18 (16) & 120.00 (2) \\ [1ex]
 \hline\\[1ex]
Mn$_{2.97}Fe_{0.30}$Ge ($\delta$=0.30) & 5.313(2)  & 5.314(2)  & 4.302(18)  & 90.06 (3)& 89.91 (3) & 119.93 (4) \\[1ex]
 \hline\\[1ex]
Mn$_{2.69}Fe_{0.62}$Ge ($\delta$=0.62) & 5.294(6) & 5.282(8) & 4.281(11)  & 90.5 (2) & 89.7 (2) & 120.00 (3) \\[1ex]
 \hline
\end{tabular*}
\label{T1}
\end{table*}

\begin{figure}[hb]
\centering
	\includegraphics[width=\linewidth]{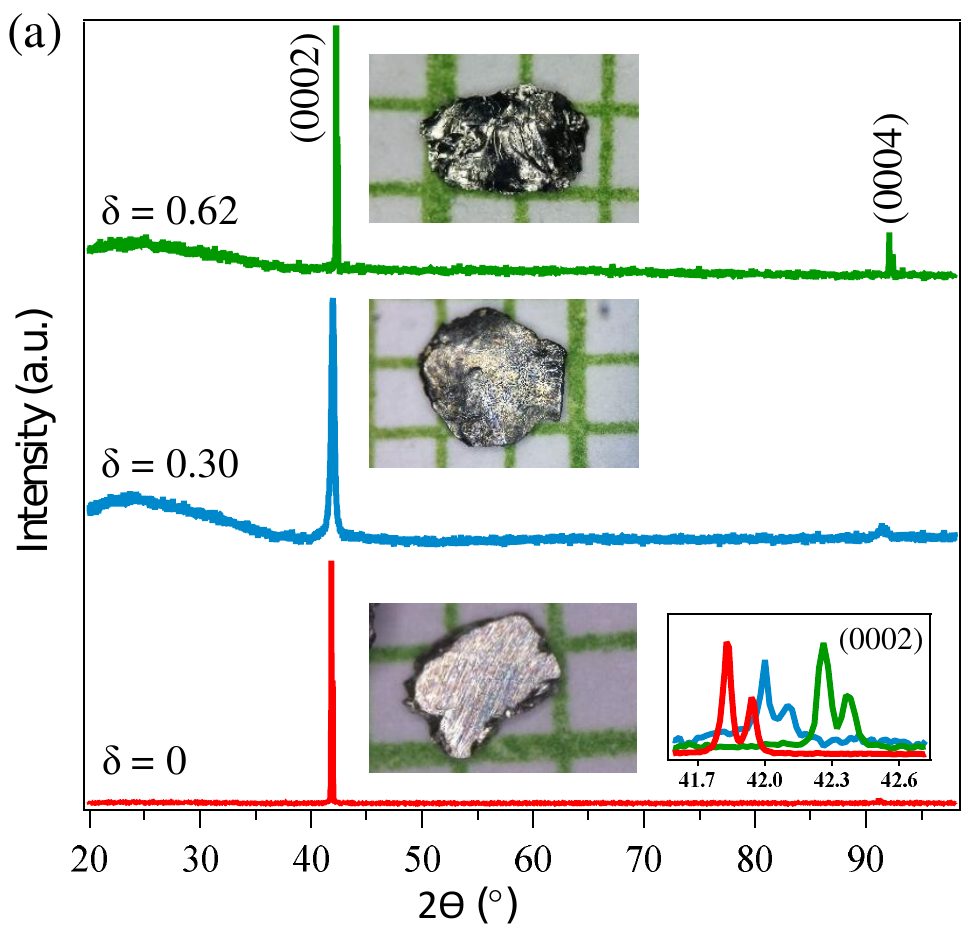}
	\caption{Powder XRD patterns of  Mn$_{3.48}$Ge ($\delta$=0), Mn$_{2.97}$Fe$_{0.30}$Ge ($\delta$=0.30), and Mn$_{2.69}$Fe$_{0.62}$Ge ($\delta$=0.62) single crystals. Bottom-right inset shows zoomed-in (0002) peaks of $\delta$=0, 0.30, and 0.62 compounds, demonstrating the shift in peak position with doping. Photographic image of the respective compositions are shown.}
	\label{1}
\end{figure}

\section{Experimental details}
Single crystals of Mn$_{(3+x)-\delta}$Fe$_{\delta}$Ge ($\delta$=0, 0.30, and 0.62) were prepared by the melt-growth method using a high-temperature muffle furnace. In this method, Manganese (Alfa Aesar 99.95\%), Iron (Alfa Aesar 99.99\%), and Germanium (Alfa Aesar 99.999\%) powders were taken in stoichiometric ratio, mixed thoroughly in the glove-box under an argon environment before sealing them in a preheated quartz ampoule under a vacuum of 10$^{–4}$ mbar. The sealed quartz tube was then heated in a muffle furnace up to $1050^{o}$C and kept at that temperature for 24 hours. The tube was slowly cooled to $760^{o}$C for $\delta$=0, $780^{o}$C for $\delta$=0.30, and $800^{o}$C for $\delta$=0.62 at a rate of 2 K/h. After prolonged annealing at the respective growth temperatures for five more days, the ampules were quenched in ice water to avoid the impurity phases. As-grown single crystals of Mn$_{(3+x)-\delta}$Fe$_{\delta}$Ge were looking shiny and had a typical size of 2$\times$1.5 mm$^2$.

Phase purity and crystal structure of the single crystals were examined by using single crystal X-ray diffraction (SCXRD, SuperNova, Rigaku) and powder X-ray diffraction (XRD, Rigaku SmartLab) with Cu-K$_\alpha$ radiation of wavelength $\lambda$=1.5406 $\AA$. Using energy dispersive X-ray spectroscopy (EDS of EDAX), we find the actual chemical composition of the as-prepared samples to be Mn$_{3.48\pm0.02}$Ge, Mn$_{2.97\pm0.05}$Fe$_{0.30\pm0.02}$Ge, and Mn$_{2.69\pm0.04}$Fe$_{0.62\pm 0.02}$Ge. Usually, Mn$_{3+x}$Ge forms with excess Mn without any control on the Mn concentration ~\cite{Yamada1988, Kiyohara2016, Xu2020, Chen2021}. Thus, we also got the crystals with excess Mn. For convinience, we represent the systems Mn$_{3.48}$Ge, Mn$_{2.97}$Fe$_{0.30}$Ge, and Mn$_{2.69}$Fe$_{0.62}$Ge in the manuscript by $\delta$=0, 0.30, and 0.62, respectively, wherever applicable. Electrical transport, Hall effect, and magnetic property studies were performed using a 9-Tesla physical property measurement system (PPMS, Dynacool, Quantum Design) within the temperature range of 2–380 K. Electrical transport and Hall measurements were performed using the four-probe technique. Copper leads were attached to the sample using EPO-TEK H20E silver epoxy.

\begin{figure*}[ht]
	\includegraphics[width=\linewidth]{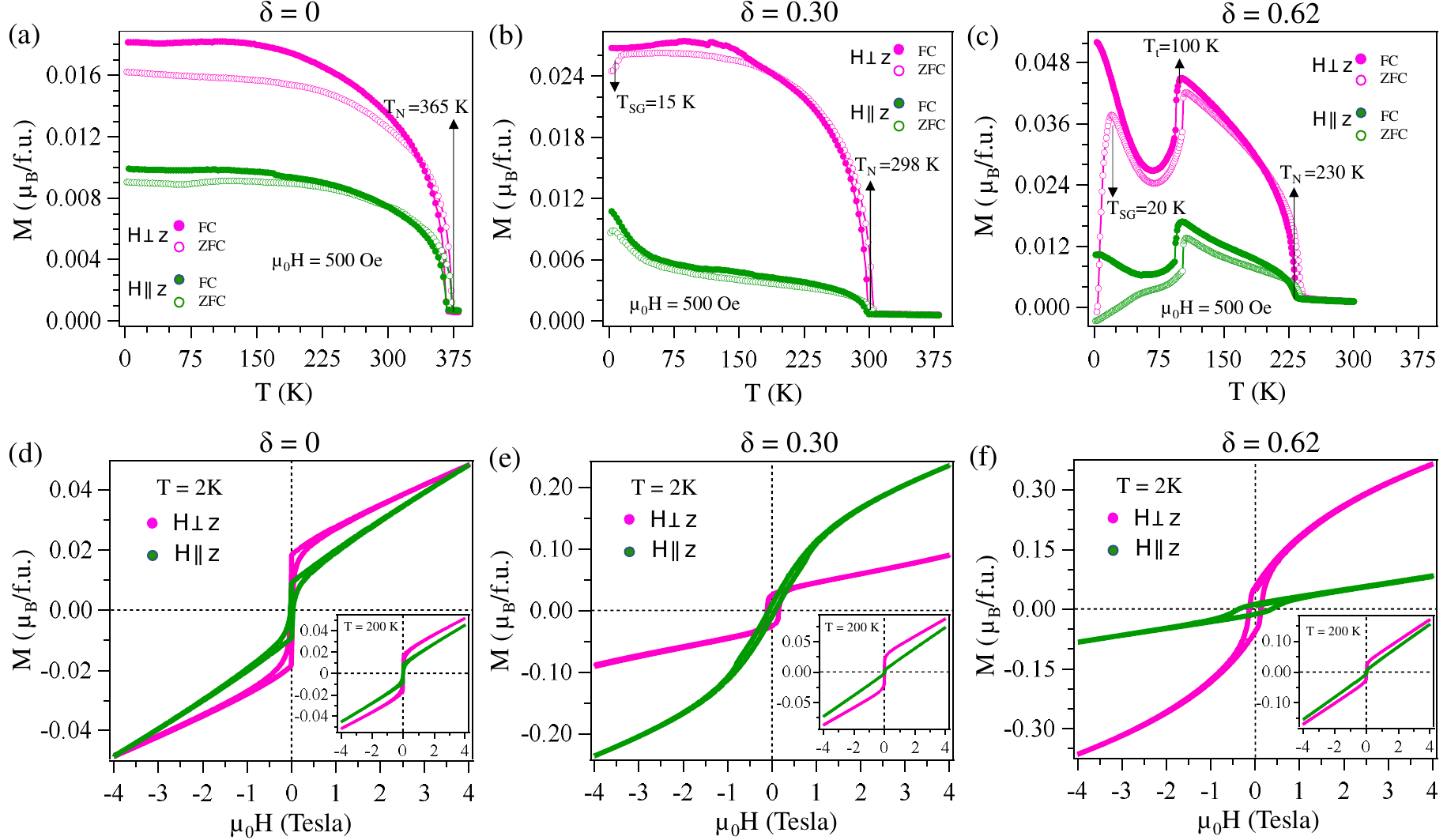}
	\caption{Temperature-dependent magnetization [$M(T)$] of Mn$_{(3+x)-\delta}Fe_{\delta}$Ge measured in zero-field-cooled (ZFC) and field-cooled (FC) modes with $H\parallel z$ and $H\perp z$ for $\delta$=0 (a), $\delta$=0.30 (b), and $\delta$=0.62 (c). Similarly, magnetization isotherms [$M(H)$] measured at 2 K with $H\parallel z$ and $H\perp z$ for $\delta$=0 (d), $\delta$=0.30 (e), and $\delta$=0.62 (f). Insets of (d), (e), and (f) are $M(H)$ data measured at 200 K from $\delta$=0, 0.30, and 0.62, respectively. See Supplemental Material for the $M(H)$ data measured at several other sample temperatures.}
	\label{fig2}
\end{figure*}

\section{Results}

\subsection{Structural Properties}
Mn$_{3}$Ge crystalizes into the Ni$_{3}$Sn-type hexagonal structure with a space group of $P6_{3}/mmc$ (194), with the Mn atoms sitting on the $ab$-plane to form a breathing kagome lattice, while the Ge atoms sit at the center of the hexagons \cite{Nayak2016, Kiyohara2016}. A couple of kagome lattice planes are stacked along the $c$-axis per unit cell. XRD patterns taken on the single crystal of Mn$_{(3+x)-\delta}Fe_{\delta}$Ge is shown in Fig.~\ref{1}, suggesting that the crystal growth is parallel to the $\it{c}$-axis. The bottom right inset in Fig.~\ref{1} shows zoomed-in (0002) peaks of the parent, $\delta$=0.30, and $\delta$=0.62 compounds. Here we observe that with increased Fe doping, (0002) peaks are shifted to higher 2$\theta$ values. This indicates a decrease in the lattice parameter $\it{c}$. Optical images of the as-grown single crystals of different compositions are shown in the middle insets of Fig. \ref{1}. Lattice parameters obtained from the single crystal XRD of $\delta$=0, 0.30, and 0.62 compounds are tabulated in Tab. \ref{T1}.

\subsection{Magnetic Properties}
To explore the magnetic properties, magnetization as a function of temperature [$M(T)$] was measured in the field-cooled (FC) and zero-field-cooled (ZFC) modes under a magnetic field ($H$) of 500 Oe applied parallel ($H\parallel z$) and perpendicular ($H\perp z$) to the $\it{z}$-axis as shown in Figs.~\ref{fig2}(a), ~\ref{fig2}(b), and ~\ref{fig2}(c) from the $\delta$=0, 0.30, and 0.62 compounds, respectively. From Fig.~\ref{fig2}(a),  we identify an antiferromagnetic (AFM) transition at a  N$\acute{e}$el temperature of T$_N$=365 K in Mn$_{3.48}$Ge, consistent with previous reports on Mn$_{3+x}$Ge where $T_N$ was found between 365 K and 400 K depending on the $\it{x}$ value~\cite{Yamada1988, Nayak2016, Kiyohara2016, Chen2021, Wuttke2019}. Below $T_N$, the magnetization increases with decreasing temperature and saturates at very low temperatures for both $H\parallel z$ and $H\perp z$ directions. From Fig.~\ref{fig2}(b), we observe a similar behavior of magnetization from the $\delta$=0.30 sample, except that $T_N$ is reduced to 298 K. A down-turn in the ZFC $M(T)$ data is noticed at a very low sample temperature of 15 K, which is clearly visible with $H\perp z$. This hints at a spin-glass-like transition induced by the Fe doping \cite{Dho2002, Feng2006, Wang2010, Bag2018}. Next, from Fig.~\ref{fig2}(c), we observe a further decrease in $T_N$ to 230 K with a Fe doping of $\delta$=0.62 in addition to a sudden drop in magnetization at 100 K from both $H\parallel z$ and $H\perp z$ directions. This kind of magnetization drop earlier reported on Mn$_3$Sn was understood as an AFM-to-AFM magnetic transition \cite{Duan2015, Sung2018}.  Moreover, at  20 K, we observe a downturn in the ZFC $M(T)$ data, clearly visible from the $H\perp z$ direction, which is sharper than $\delta$=0.30 compound.

Interestingly, from the $M(T)$ of Mn$_{3.48}$Ge measured in the FC mode, we notice almost two times higher in-plane magnetization ($H\perp z$) than the out-of-plane magnetization ($H\parallel z$) at 2 K, indicating a strong magnetic anisotropy in this systems. Such strong magnetic anisotropy in Mn$_{3.48}$Ge, despite being an AFM metal, mainly originates from the geometrical frustration of the Mn magnetic moments within the kagome lattice plane, producing a finite net magnetic moment \cite{Yamada1988, Tomiyoshi1983}. Further, from Figs.~\ref{fig2}(b) and ~\ref{fig2}(c) we can notice that the in-plane magnetization increases nearly three times and four times compared to the out-of-plane magnetization for the Fe doping concentration of $\delta$=0.30 and 0.62, respectively. Thus, the magnetic anisotropy rapidly increases with Fe doping concentration.

\begin{figure}[ht]
\centering
\includegraphics[width=\linewidth, clip=true]{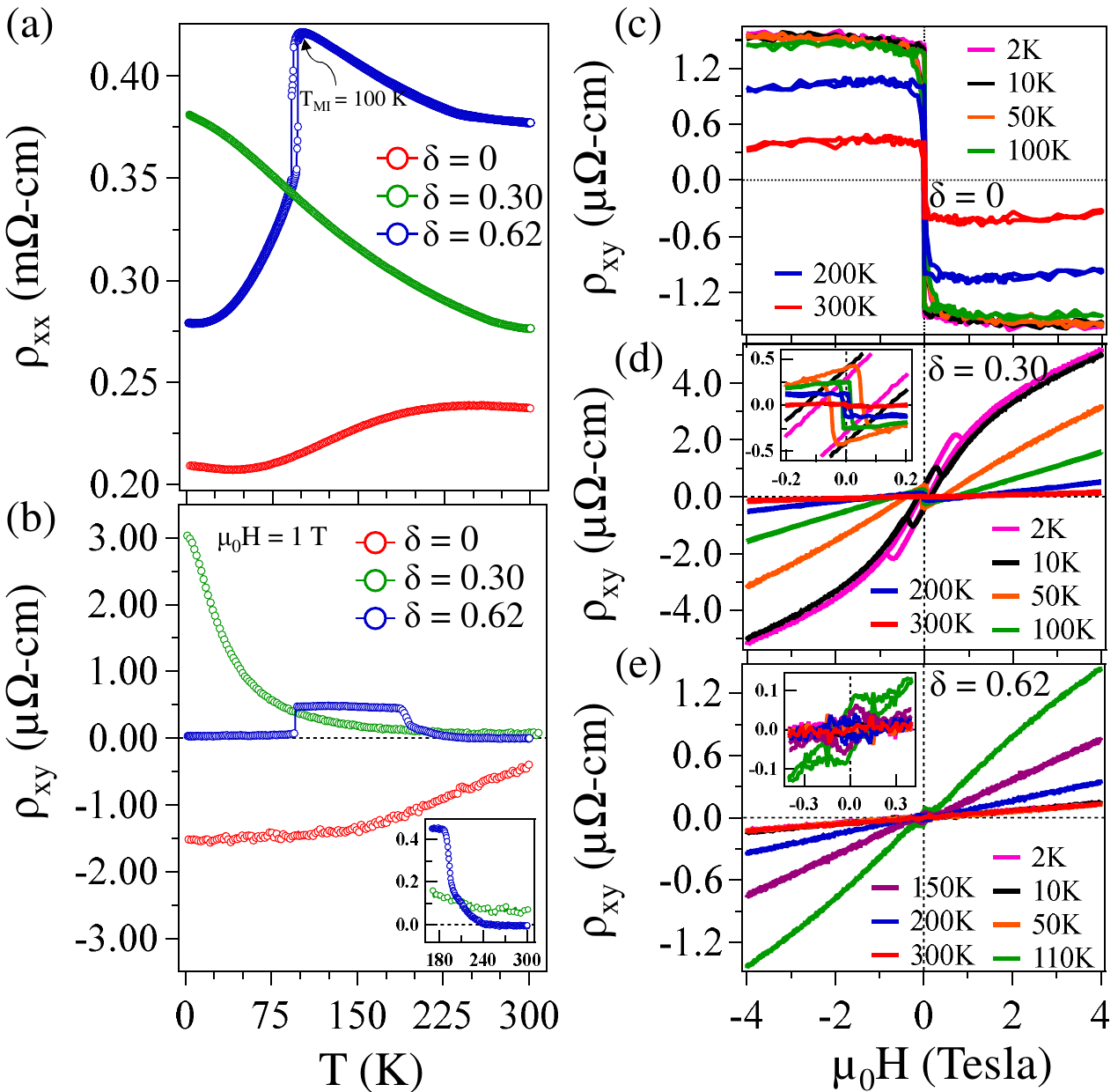}
\caption{Temperature-dependent (a) longitudinal resistivity ($\rho_{xx}$) and (b) Hall resistivity ($\rho_{xy}$) plotted for $\delta$=0, 0.30, and 0.62. Inset in (b) is the zoomed-in image taken to demonstrate the increase in total Hall resistivity of $\delta$=0.62 below 230 K.   Field-dependent Hall resistivity ($\rho_{xy}$) measured at various temperatures for $\delta$=0 (c), $\delta$=0.30 (d), and $\delta$=0.62 (e) compounds. Inset in (d) and (e) are the zoomed-in images taken around zero field.}
\label{fig4}
\end{figure}

Figs.~\ref{fig2}(d), \ref{fig2}(e), and \ref{fig2}(f) show magnetization isotherms [$M(H)$] measured at 2 K with $H\parallel z$ and $H\perp z$ directions from the $\delta$=0, 0.30, and 0.62 compounds, respectively. From Fig.~\ref{fig2}(d), we observe a spontaneous magnetization at lower fields, which then linearly increases with the field. Further, consistent with $M(T)$ data [see Fig.~\ref{fig2}(a)], the in-plane spontaneous magnetization is nearly two times higher than that of out-of-plane magnetization at lower fields. Inset in Fig.~\ref{fig2}(d) shows $M(H)$ data taken at 200 K. Nevertheless, we do not find any significant difference in the $M(H)$ data between 2 and 200 K. On the other hand, more interestingly, from the $M(H)$ data of the $\delta$=0.30 compound [see Fig.~\ref{fig2}(e)], we find a quite complex magnetization isotherm. That means the out-of-plane ($H\parallel z$) magnetization increases more rapidly than the in-plane ($H\perp z$) magnetization with increasing applied field and eventually dominates the in-plane magnetization beyond 0.3 T of applied field when measured at 2 K. We further notice that at 4 T of the applied field, the out-of-plane magnetization is nearly three times higher than that of the in-plane. This observation is in stark contrast to the parent system, where we observe dominating in-plane magnetization throughout the applied field up to 4 T. The inset in Fig.~\ref{fig2}(e) shows the isotherm measured at 200 K, demonstrating that the in-plane magnetization dominates the out-of-plane throughout the applied field up to 4 T, which is similar to the parent system. Further, Fe doping of $\delta$=0.62 converts the system more into an in-plane ferromagnet (with a sigmoid-like $M(H)$ loop) and an out-of-plane antiferromagnet (with a linear $M(H)$ loop), as shown in Fig.~\ref{fig2}(f).

\subsection{Electrical and Magnetotransport Properties}

Fig.~\ref{fig4}(a) depicts the zero-field in-plane longitudinal electrical resistivity ($\rho_{xx}$) of the $\delta$=0, 0.30, and 0.62 compounds measured between 2 and 300 K. We notice from the resistivity that the parent compound, Mn$_{3.48}Ge$, exhibits overall a metallic behavior with temperature that is in agreement with earlier reports on Mn$_{3+x}$Ge~\cite{Kiyohara2016}. On the other hand, with the Fe doping of $\delta$=0.30,  the system exhibits a semiconducting or a bad-metallic behavior as $\rho_{xx}$  increases with decreasing temperature. With further increasing the Fe doping to $\delta$=0.62, we observe that the $\rho_{xx}$ increases with decreasing temperature before a sharp drop in the resistivity takes place at 100 K [see Fig.~\ref{fig4}(a)], below 100 K the system behaves like a metal. Such a sudden drop in the resistivity at 100 K is possibly due to the in-plane AFM-to-AFM magnetic transition [see Fig.~\ref{fig2}(c)] leading to the metal-insulator (MI) transition as reported earlier on  Mn$_{3-\delta}$Fe$_{\delta}$Sn~\cite{Low2022} and on Mn$_{2.34}$Fe$_{0.66}$Ge~\cite{Rai2023}.

\begin{figure*}[t]
\centering
	\includegraphics[width=\linewidth]{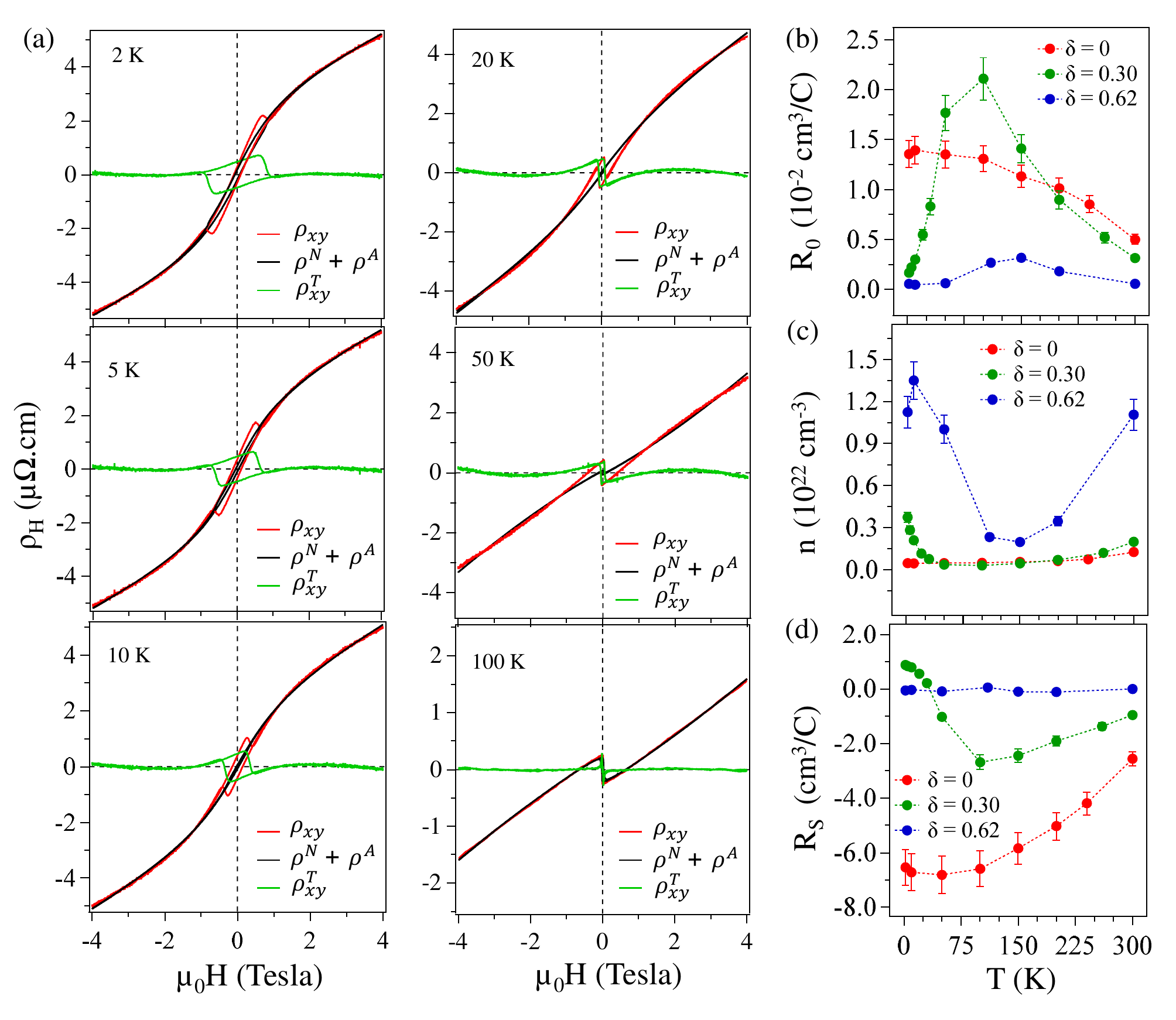}
	\caption{(a) Field-dependent in-plane Hall resistivity ($\rho_{\it{xy}}$) measured at different temperature of $\delta$=0.30. The red curves are experimental data and the solid-black curves are the fits using the Eqn.~\ref{eq1} and green curves are the topological Hall resistivity  (${\rho}^T_{xy}$). (b) Normal Hall coefficient ($R_0$). (c) Carrier density plotted as a function of temperature. (d) Anomalous Hall coefficient ($R_S$) plotted as a function of temperature.}
	\label{fig5}
\end{figure*}

Fig.~\ref{fig4}(b) depicts the Hall resistivity [$\rho_{\it{xy}}(T)$] $\delta$=0, 0.30, and 0.62 compounds measured as a function of temperature. Here, $\rho_{\it{xy}}$ stands for the in-plane Hall resistivity ($\it{ab}$-plane) measured with the current applied parallel to the $\it{x}$-axis and field applied parallel to the $\it{z}$-axis, while the Hall voltage is measured along the $\it{y}$-axis. Hall resistivity was measured using both positive (+H) and negative (-H) fields to exclude the magnetoresistance contribution due to possible misalignment of the four-probe connection. The resultant Hall resistivity was calculated using the formula, $\frac{\rho_H (H,T)-\rho_H (-H,T)}{2}$. Thus, from Fig. \ref{fig4}(b), we can observe that the $\rho_{xy}$ of the parent compound monotonically increases (negative value) with decreasing temperature which then saturates at low temperatures. This observation agrees with previous reports on this system \cite{Nayak2016}. $\rho_{xy}$(T) of the parent system bears a resemblance to the $M(T)$ data [see Fig.~\ref{fig2}(a)]. On the other hand, the $\rho_{xy}$ of $\delta$=0.30 compound rapidly increases (positive) with decreasing temperature, possibly due to a rapid increase in the out-of-plane magnetization [see Fig.~\ref{fig6}(d)]. Next, from the $\rho_{xy}$ of $\delta$=0.62, we can find significant Hall signal only within the temperature range of 100–230 K, which is consistent with the $M(T)$ data as shown in Fig.~\ref{fig2}(c) where we notice one antiferromagnetic transition (AFM) at $T_{N}$=230 K and the other one (also AFM type) at $T_{t}$=100 K.

Field-dependent Hall resistivity [$\rho_{xy}(H)$] measured at various temperatures is plotted in Figs.~\ref{fig4}(c), ~\ref{fig4}(d), and ~\ref{fig4}(e) from the $\delta$=0, 0.30, and 0.62 compounds, respectively. From the parent compound, we observe a sudden jump in $\rho_{xy}$ near the zero fields, which saturates with further increasing the field as shown in Fig.~\ref{fig4}(c), which is a signature of the anomalous Hall effect (AHE). This observation is consistent with previous reports on similar systems \cite{Nayak2016, Kiyohara2016}. Also, in agreement with $\rho_{xy}(T)$ as shown in Fig.~\ref{fig4}(b), the saturated $\rho_{xy}(H)$ value increases with decreasing temperature and becomes nearly constant below 100 K. On the other hand, with a Fe doping of $\delta$=0.30, the AHE is significantly decreased, as we do not observe a sharp jump near the zero fields but only a gradual increase in $\rho_{xy}$ with the field. Particularly at higher temperatures ($>$ 100 K), we observe a linear dependence of $\rho_{xy}$ on the applied field due to the dominating normal Hall contribution. However, at lower temperatures ($<$ 100 K), we still find a non-linear and hysteretic $\rho_{xy}(H)$ behavior. Further increasing the Fe doping to $\delta$=0.62, as shown in Fig.~\ref{fig4}(e), the AHE signal is hardly visible at most of the measured temperatures except a small AHE is noticed at 110 and 200 K.   Thus, the total Hall resistivity of $\delta$=0.62 shown in Fig.~\ref{fig4}(b) is mainly contributed by the normal Hall resistivity except for the temperature range of 100-230 K where a small AHE contribution also present.

Red-colored data in Fig.~\ref{fig5}(a) depict the total Hall resistivity of $\delta$=0.30 fitted using Eqn.~\ref{eq1} measured at various sample temperatures. Black-colored data in Fig.~\ref{fig5}(a) are the fits using the Eqn.~\ref{eq1}. We can notice that $\rho_{xy}(H)$ is not fitted well at lower fields because of the topological Hall contribution. Therefore, to extract the topological Hall contribution, one needs to subtract anomalous and normal Hall contributions from the total Hall resistivity such as $\rho^T_{xy}(H)$ = $\rho_{xy}(H)$-[${\rho}^N_{xy}(H)$ + ${\rho}^A_{xy}(H)$]. Green-colored data in Fig.~\ref{fig5}(a) depict the obtained topological Hall resistivity plotted as a function of field at various temperatures. From Fig.~\ref{fig5}(a), we notice that the topological Hall effect is more significant at low temperatures with a maximum of $\rho^T_{xy}$ = 0.69 $\mu\Omega$ cm for a critical field of 0.6 T at 2 K, which disappears at higher temperatures ($>$50 K).

\begin{equation}\label{eq1}
\rho_H = {\rho}^N_H + {\rho}^A_H = R_0\mu_0H + R_S\mu_0M
\end{equation}
Here, R$_0$ is the normal Hall coefficient and R$_S$ is the anomalous Hall coefficient.

Next, in addition to $\rho_{xy}(H)$ of $\delta$=0.30, $\rho_{xy}(H)$ of $\delta$=0 and 0.62 were also fitted by the Eqn. \ref{eq1} to estimate the values of normal (R$_0$) and anomalous (R$_S$) Hall coefficients. Fig. \ref{fig5}(b) shows temperature-dependent normal Hall coefficient. Since the normal Hall coefficient R$_0$ is related to carrier density ($n$) and charge ($q$) as $R_0=\frac{1}{n|q|}$, we can obtain the carrier concentration as a function of temperature,  shown in Fig. \ref{fig5}(c). From Fig. \ref{fig5}(c), we can notice that the carrier concentration of the parent system is almost constant with temperature. For $\delta$=0.30, it hardly changes with temperature down to 50 K and with further decreasing the temperature the carrier concentration rapidly increases.  On the other hand, the carrier concentration of $\delta$=0.62 is nearly the same at both 2 and 300 K but it significantly decreases as the sample approaches 100 K. Such a change in carrier density is possibly related to the magnetic transition around this temperature. Further, the anomalous Hall effect is quantified by the anomalous Hall coefficient (R$_S$). From Fig. \ref{fig5}(d), we can find large R$_S$ values for the parent system at 2 K, which decrease with increasing temperature, consistent with the total Hall resistivity [see Fig. \ref{fig4}(b)]. On the other hand, fluctuating R$_S$ values are noticed in the $\delta$=0.30 compound. To be precise, positive and negative $R_S$ values are found below and above the sample temperature of 50 K, respectively. Such a crossover from positive to negative $R_S$ values indicates the sign change in the anomalous Hall effect as observed in the inset of Fig.~\ref{fig4}(d)~\cite{Gu2022}. Negligible R$_S$ values found from the $\delta$=0.62 suggest negligible AHE which is in agreement with Fig.~\ref{fig4}(e). We further notice a change in the sign of AHE between the parent system ($-ve$)  and the Fe doped system of $\delta$=0.62 ($+ve$), possibly due to the change in carrier concentration with Fe doping as a sign of the Hall conductivity or Berry curvature is sensitive to the chemical potential~\cite{Xu2020, Wang2021}. The same can be valid for the $\delta$=0.30 system as well, in which we observe a change in the sign of AHE below ($+ve$) and above ($-ve$) 50 K [see Figs.~\ref{fig5}(c) and \ref{fig5}(d)].

\section{Discussions}
Overall, the electrical transport, magnetic, and magnetotransport properties of Mn$_{3.48}$Ge are consistent with previous reports \cite{Yamada1988, Nayak2016, Kiyohara2016, Xu2020, Chen2021}. During this manuscript preparation, a report on Mn$_{2.34}$Fe$_{0.66}$Ge has appeared in the literature \cite{Rai2023}, thoroughly discussing the magnetic and magnetotransport properties. Our magnetic and magnetotransport results on $\delta$=0.62 compound agree with Ref.~\cite{Rai2023}. That means the antiferromagnetic transitions observed at 230 K and 100 K in our studied system are very close to the reported values of 240 K and 120 K, respectively. Further, the absence of in-plane anomalous Hall effect ($\rho_{xy}$) is consistent with the report. However, a small out-of-plane AHE  ($\rho_{zy}$) is reported in Ref.~\cite{Rai2023} with a maximum of $\rho_{zy}$=0.7 $\mu\Omega$ cm at 130 K, which we could not study on our skinny samples. Nevertheless, the dominating normal Hall resistivity at all measured temperatures is in excellent agreement with Ref.~\cite{Rai2023}. In addition, the electrical transport and magnetic properties of our studied system Mn$_{2.69}$Fe$_{0.62}$Ge are consistent with several other previous reports on Mn$_{3-\delta}$Fe$_{\delta}$Ge ($\delta\geq0.45$)~\cite{Hori1992, Hori1995, Du2007}. Specifically, the metal-insulator transition observed at 100 K in our $\delta$=0.62 system is in agrement with the previous report on (Mn$_{0.83}$Fe$_{0.2}$)$_{3.25}$Ge~\cite{Du2007}.

Next, the new results of this manuscript, not yet reported so far, are the electrical transport, magnetic, and Hall effect on Mn$_{2.97}$Fe$_{0.30}$Ge compound. Interestingly, although both the parent and $\delta$=0.62 compounds show an increase in longitudinal resistivity ($\rho_{xx}$) with temperature, at least in the low-temperature regime, Mn$_{2.97}$Fe$_{0.30}$Ge does not display any metallic nature of the resistivity. Means the highest resistivity observed at 2 K gradually decreases with increasing temperature, like in magnetic semiconductors \cite{Dietl2010, Deng2011, Yu2020}. On the other hand, we could observe the topological Hall effect in $\delta$=0.30 which was not found in the parent and  $\delta$=0.62 systems. There exist several mechanisms to understand the origin of the topological Hall effect, such as the Dzyaloshinskii-Moria (DM) interaction in noncentrosymmetric systems \cite{Kanazawa2015,Liu2017a,Sukhanov2020},  uniaxial magnetocrystalline anisotropy in centrosymmetric systems \cite{Yu2014,Preissinger2021,Low2022}, and chiral domain-wall-induced skyrmion lattice \cite{Gudnason2014,Cheng2019,Nagase2021,Yang2021}. Furthermore, the recent theoretical and experimental studies suggest the stabilization of skyrmion lattice in centrosymmetric kagome systems~\cite{Okubo2012,Hayami2017,Rout2019} such as $Gd_2PdSi_3$~\cite{Kurumaji2019}, $Gd_3Ru_4Al_{12}$~\cite{Hirschberger2019} due to  geometrical magnetic frustration. Recently, we have also shown an enhancement of magnetocrystalline anisotropy in addition to the formation of the noncoplanar spin structure generating the topological Hall effect in Mn$_{3-\delta}$Fe$_{\delta}$Sn \cite{Low2022}. The same could be true in the present case of $\delta$=0.30 compound as well.

\begin{figure}[t]
\centering
	\includegraphics[width=\linewidth]{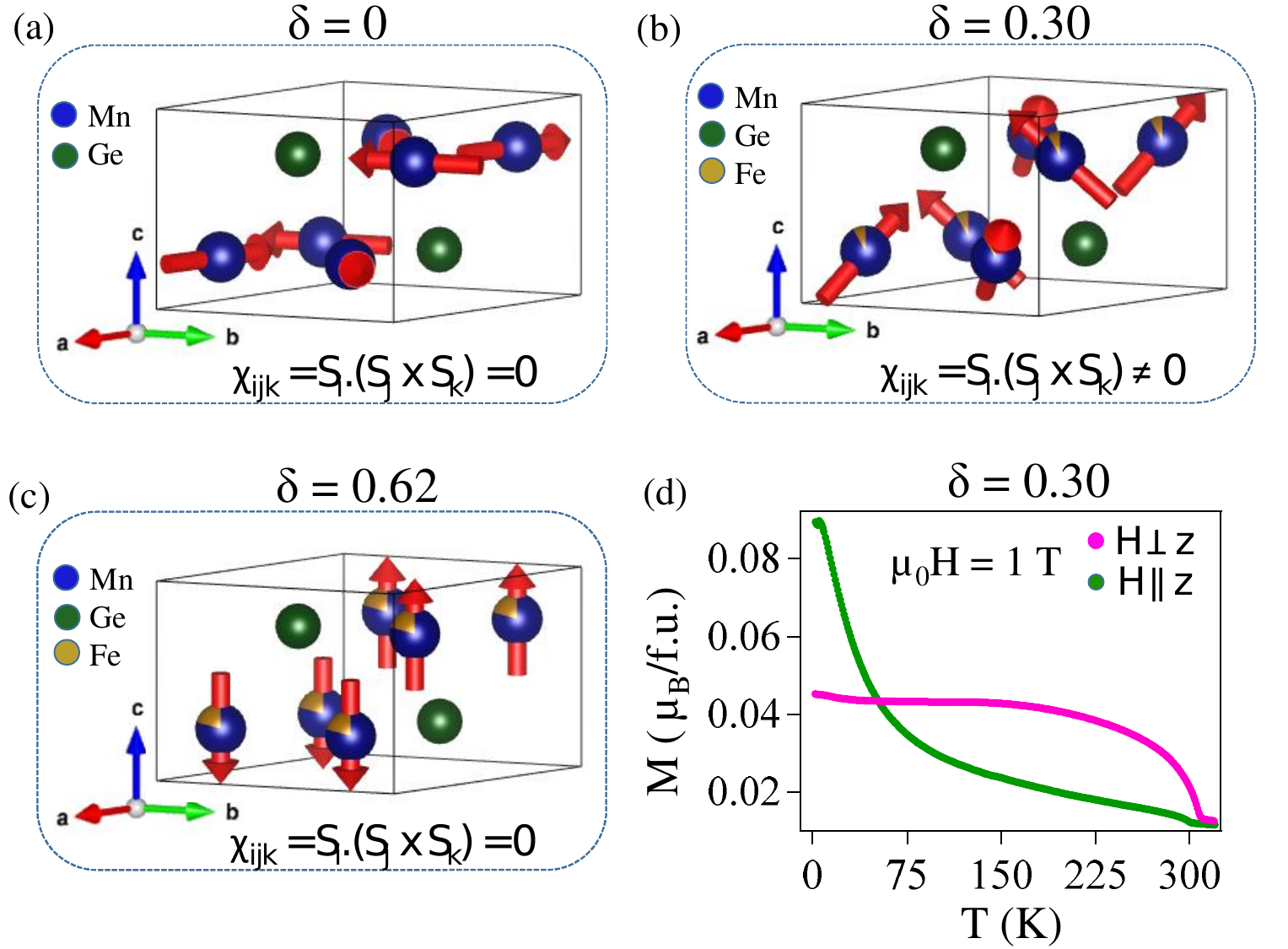}
	\caption{Schematic representation of the Mn magnetic moment arrangement in the parent (a), $\delta$=0.30 (T $<$ 50 K) (b), and $\delta$=0.62 (T $<$ 100 K) (c) compounds. (d) Magnetization plotted as a function of temperature measured at 1 T of applied field from $\delta$=0.30 sample.}
	\label{fig6}
\end{figure}

Fig.~\ref{fig6} schematically demonstrates the magnetic structure observed in the studied compositions of $\delta$=0, 0.30, and 0.62. Fig.~\ref{fig6}(a) depicts the Mn$_3$Ge magnetic structure with an inverse-triangular arrangement of the Mn magnetic moments. Here, the Mn magnetic moments are in a noncollinear but coplanar antiferromagnetic order,  leading to zero scalar spin chirality $\chi_{ijk}= S_i.(S_j \times S_k) =$ 0. Thus, no topological Hall effect has been observed but only the anomalous Hall effect due to nonzero Berry curvature in the momentum space~\cite{Chen2021, Nakatsuji2015, Kiyohara2016, Nayak2016}.  With a Fe doping of $\delta$=0.62, Mn$_{2.69}$Fe$_{0.62}$Ge has a collinear out-of-plane AFM ordering as demonstrated in Ref.~\cite{Rai2023} by the neutron diffraction study as well in this study [see Fig.~\ref{fig2}(f)],  is schematically shown in Fig.~\ref{fig6}(c). Since in this case also the scalar spin chirality is zero, no topological Hall effect is anticipated.


Fig.~\ref{fig6}(b) schematically shows the noncoplanar arrangement of the Mn magnetic moments in Mn$_{2.97}$Fe$_{0.30}$Ge, where the magnetic moments align neither ferromagnetically nor antiferromagnetically in any particular direction under the external applied field. In this case, the scalar spin chirality will be nonzero, $\chi_{ijk}= S_i.(S_j \times S_k) \neq0$. As a result, the topological Hall resistivity is warranted, as shown in Fig.~\ref{fig5}(a). Further, to confirm that the system has a noncoplanar spin structure, we plotted $M(T)$ of Mn$_{2.97}$Fe$_{0.30}$Ge under the applied field of 1 T in Fig.~\ref{fig6}(d). From Fig.~\ref{fig6}(d), we can notice that the in-plane magnetization gradually increases with decreasing temperature and nearly saturates below 150 K. On the other hand, the out-of-plane magnetization rapidly increases with decreasing temperature and dominates the in-plane magnetization below 50 K. Such a crossover from dominating in-plane to out-of-plane magnetization indicates that the spins are canted towards out-of-plane directions, producing field-induced noncoplanar spin structure. This argument is further supported by Fig.~\ref{fig2}(e), in which the in-plane magnetization dominates the out-of-plane magnetization at higher temperatures (T = 200 K). In contrast, the out-of-plane magnetization dominates the in-plane at 2 K, above 0.3 T of applied fields. Therefore, our results suggest that the noncoplanar spin structure stabilized by the strong magnetocrystalline anisotropy originates the topological Hall effect in  Mn$_{2.97}$Fe$_{0.30}$Ge~\cite{Low2022}.


\section{Summary }

We have systematically studied the electrical, magnetic, and magnetotransport properties of Mn$_{(3+x)-\delta}Fe_{\delta}Ge$ ($\delta$=0, 0.30, and 0.62). We find that the electrical resistivity of the parent compound displays metallic behavior, while the system with $\delta$=0.30 of Fe doping exhibits resistivity similar to a dilute magnetic semiconductor. With further Fe doping of $\delta$=0.62, the system demonstrates a metal-insulator transition at 100 K. Fe doping increases ferromagnetism and magnetocrystalline anisotropy. It induces a spin-glass-like state at low temperatures. In addition, the spontaneous anomalous Hall effect observed in the parent system is significantly reduced with increasing Fe doping concentration. Importantly, the topological Hall effect observed in Mn$_{2.97}$Fe$_{0.30}$Ge ($\delta$=0.30),  is not found from the parent system Mn$_{3.48}$Ge or Mn$_{2.69}$Fe$_{0.62}$Ge ($\delta$=0.62).

\section{Acknowledgement}

S.G. acknowledges the University Grants Commission (UGC), India, for the Ph.D. fellowship. The authors thank SERB (DST), India, for the financial support (Grant No. SRG/2020/00393). This research has used the Technical Research Centre (TRC) Instrument facilities of the S. N. Bose National Centre for Basic Sciences, established under the TRC project of the Department of Science and Technology (DST), Govt. of India. The authors thank Prof. Kamaraju Natarajan and Siddhanta Sahu, IISER Kolkata, for the SCXRD measurements.

\section*{References}
\bibliographystyle{iopart-num}
\bibliography{Mn3Ge}

\end{document}